\documentclass[pra,twocolumn,superscriptaddress,floatfix,amsmath,nofootinbib,amssymb]{revtex4-1}
\usepackage[UKenglish]{babel}
\usepackage{graphicx}% Include figure files
\usepackage{dcolumn}% Align table columns on decimal point
\usepackage{bm}% bold math
\usepackage{verbatim}
\usepackage{mathrsfs}
\usepackage{color}
\newcommand{\ket}[1]{\left| {#1} \right\rangle}

\newcommand{\braket}[2]{\left\langle {#1}\left|{#2}\right.\right\rangle}
\newcommand{\proj}[2]{\left| {#1} \right\rangle\!\left\langle {#2} \right|}

\newcommand{\ro}{r_{\omega}}
\newcommand{\eq}[1]{(\ref{#1})}
\newcommand{\sgn}{\text{sgn}}
\newcommand{\up}{\uparrow}
\newcommand{\down}{\downarrow}
\newcommand{\qr}{q_\text{R}}
\newcommand{\ql}{q_\text{L}}

% PACS, the Physics and Astronomy
\begin{document}

\title{Fermionic entanglement ambiguity in non-inertial frames}
\author{Miguel Montero}
\affiliation{Instituto de F\'{i}sica Fundamental, CSIC, Serrano 113-B, 28006 Madrid, Spain}
\author{Eduardo Mart\'{i}n-Mart\'{i}nez}
\affiliation{Instituto de F\'{i}sica Fundamental, CSIC, Serrano 113-B, 28006 Madrid, Spain}

\begin{abstract}
We analyse an ambiguity in previous works on entanglement of fermionic fields in non-inertial frames. This ambiguity, related to the anticommutation properties of field operators, leads to nonunique results when computing entanglement measures for the same state. We show that the ambiguity disappears when we introduce detectors, which are in any case necessary as a means to probe the field entanglement. 
\end{abstract}

\maketitle

\section{Introduction}

We will discuss that for bipartite systems of fermionic fields, anticommutation of fermionic operators naturally induces an ambiguity when defining the individual basis of the observers. This ambiguity is related to the ordering criterion of the creation and annihilation operators and has gone unnoticed in all previous literature on fermionic field entanglement in non-inertial frames \cite{AlsingSchul,Edu2,Edu3,chapucilla,chapucilla2,Shapoor,Geneferm,chor1,Edu9,chor2}.

Such ordering is only a mathematical convention since, obviously, physical states are the same regardless of how we rearrange the fermionic operators. Therefore, one would expect that choosing a different convention should not change the behaviour of entanglement measures. We will show that using the formalism of all the previous literature, the results do depend, in general, on this convention.

However, to eventually measure field entanglement, we need to consider detectors coupled to the field. We argue below that the ambiguity is removed when we incorporate detectors to the field entanglement settings. 

Incidentally we show that the convention that has a physical meaning in terms of detectors entanglement is different to that used in the previous literature, where field entanglement (without detectors) was analysed in the Dirac case \cite{Edu2} and Grassman fields beyond the single mode approximation \cite{Edu9,Edu10}.

The paper is organised as follows: In section \ref{toy} we present and thoroughly analyse the ambiguity that appears when defining bases from different orderings of fermionic operators. In section \ref{main} we study the implications of this ambiguity in the specific context of relativistic quantum information, finding that previous literature results are not independent of different sign conventions. In section \ref{ord} we show how for any physically meaningful experiment such ambiguity disappears. We suggest a physical criterion to choose the signature of the fermionic basis, based on meaningful outcomes of detectors experiments. Finally, conclusions are presented in section  \ref{cojoniak}.

\section{A toy model}\label{toy}
In this section we study the entanglement properties of a simple fermionic system to illustrate the peculiarities we intend to show in more general settings. Namely, we consider a two-mode fermionic system with fermionic creation operators $a^\dagger$ and $b^\dagger$, and a vacuum state $\ket{0}$. The relevant Hilbert space $\mathcal{H}$ is thus four dimensional. Any fermionic system in which there are only two possible one-particle states will fit to this quite general model. 

We denote the Hilbert space basis of our toy model as 
\begin{align}\ket{00}&=\ket{0},\; \ket{10}=a^\dagger\ket{0},\;\ket{01}=b^\dagger\ket{0},\nonumber\\\ket{11}&=a^\dagger b^\dagger\ket{0}.\label{bt1}\end{align}
By choosing this basis, we implicitly endow the Hilbert space with a tensor product structure which allows us to regard it as two qubits, with the first label corresponding to one qubit and the second to the other. We can now perform quantum information and study entanglement properties in this system. 

Of course, if we make a nonlocal change of basis, the entanglement of the state may change \cite{Caban}. Due to the fermionic nature of our system, there is one such change of basis which arises naturally: we can interchange the positions of $a^\dagger$ and $b^\dagger$ in \eq{bt1}, or equivalently, we may change the last term by a sign. The new basis, which we denote by a prime, is
\begin{align}
\ket{00}'&=\ket{00},\; \ket{01}'=\ket{01}\nonumber\\\ket{10}'&=\ket{10},\;\ket{11}'=b^\dagger a^\dagger\ket{0}=-\ket{11}.\label{bt2}
\end{align}
Thus, in this specific case, reversing the order of $a^\dagger$ and $b^\dagger$ leads naturally to a new basis, which only differs by a sign in one of its elements from \eq{bt1}. We usually think of the modes $a^\dagger$ and $b^\dagger$ as each one acting on one qubit. A naive expectation would be that the ordering between them should be purely conventional and should not affect entanglement. Nevertheless, a great deal changes from \eq{bt1} to \eq{bt2}: The new basis endows the Hilbert space with a different tensor product structure. By this we mean the (quite obvious) fact that the set of elements of the Hibert space which are expressible in the form $v_1\otimes v_2$, with $v_1$, $v_2$ of the form $c_1\ket{0}+c_2\ket{1}$, is different if we take the basis \eq{bt1} or \eq{bt2}. In the language of quantum information, the transformation from \eq{bt1} to \eq{bt2} is not local, and therefore it cannot be expressed as the product of two unitary transformations, with each acting on one qubit. This means that, for any given state, entanglement properties will vary wildly in general, depending on which operator ordering is taken.

We remark that this is true for any nonlocal change of basis in any bipartite quantum system. Indeed, we could have imposed the change of basis from \eq{bt1} to \eq{bt2} in a bosonic system as well. Nevertheless, whereas in a bosonic setting such a change would have been arbitrary, in the fermionic case, due to the sign acquired by operator interchange, there is no \emph{a priori} way to choose between these two bases to define our bipartite system, unless a specific set of observables with a specific operator ordering structure is chosen (see section \ref{ord}).  

As a concrete example, consider the separable state in basis \eq{bt1},
\begin{align}\ket{\Psi}&=\frac{1}{2}\left(\ket{00}+\ket{01}+\ket{10}+\ket{11}\right)\nonumber\\&=\left[\frac{1}{\sqrt{2}}\left(\ket{0}+\ket{1}\right)\right]\otimes\left[\frac{1}{\sqrt{2}}\left(\ket{0}+\ket{1}\right)\right].\end{align}
When expressed in basis \eq{bt2}, it reads
\begin{align}\ket{\Psi}&=\frac{1}{2}\left(\ket{00}'+\ket{01}'+\ket{10}'-\ket{11}'\right),\end{align}
which is a maximally entangled state as measured by the Von Neumann entropy of its reduced density matrix, which is $S=1$. In fact, if we perform the change of basis $\ket{0'}=\tfrac{1}{\sqrt{2}}\left(\ket{0}-\ket{1}\right)$, $\ket{1'}=\tfrac{1}{\sqrt{2}}\left(\ket{0}+\ket{1}\right)$ in the second qubit space, the state $\ket{\Psi}$ becomes the Bell state $\ket{\Psi^-}$.

This shows that when studying quantum information in fermionic systems it is fundamental to choose a specific operator ordering. This phenomenon happens in any fermionic system regardless of its origin, and in particular in quantum field theory of the Grassman scalar and Dirac fields, which is a point that has been neglected in all previous works known to us \cite{AlsingSchul,Edu2,Edu3,chapucilla,chapucilla2,Shapoor,Geneferm,chor1,Edu9,chor2}. 

So far, we have only described the natural consequences of making a nonlocal change of basis in a bipartite system. It could be argued that as long as we stick to one basis, albeit arbitrary, we will obtain well-defined results. However, as we show below, entanglement can also change when partial traces are taken, even if we do not change the operator ordering in the remaining state. Thus, the `ordering convention' in the unobserved degrees of freedom may seem to affect the entanglement behavior of the rest of the system.

 Consider a tripartite system. If for any reason we do not have access to the third subsystem, i.e. all of the observables that we consider belong only to the first and second subsystems, then we have to sum over all of the states of the unobserved Hilbert space. Formally, this is done by taking a partial trace over the third subsystem. After tracing, we end up with a bipartite state which is generally nonpure. 
 
In the absence of a physical criterion to adhere to a particular `operator ordering',  there is no reason to restrict ourselves to a fixed Hilbert space basis. Therefore,  we will choose the natural basis associated with each operator ordering in the same way as we did in equations \eqref{bt1} and \eqref{bt2} for the orderings $a^\dagger b^\dagger$ and $b^\dagger a^\dagger$, respectively . From now on, when we speak about `different orderings', we are implying that our sign convention when defining the Fock space basis is adapted to that ordering.
 
 We will now show that even if the untraced modes do not change their relative ordering, entanglement may change depending on the `ordering convention' in the third mode. In other words, entanglement is dependent on the relative position of traced out modes, which are no longer present after the partial trace is taken.

We will show this phenomenon by minimally modifying our toy model: Let us add a third mode created by $c^\dagger$, and consider the state
\begin{align}\ket{\Phi}=\frac{1}{2}\left(\ket{100}+\ket{010}+\ket{101}+\ket{011}\right)\label{emix}\end{align}
where the basis is defined using the ordering $(a^\dagger b^\dagger c^\dagger)$, so as to have $\ket{111}=a^\dagger b^\dagger c^\dagger\ket{0}$. Suppose that the $c^\dagger$ modes are not observed and thus results are obtained after tracing over them. This means tracing over the third label in \eq{emix}. To study entanglement between the first two subsystems we have to take a partial trace over the third one. We then study the negativity \cite{Negat} of the reduced mixed state, which would contain only $a^\dagger$ and $b^\dagger$ excitations. The negativity for state \eq{emix} after tracing out the $c^\dagger$ mode is $\tfrac{1}{2}$, so the state is entangled.

Of course, we expect that interchanging the ordering of $a^\dagger$ and $b^\dagger$ will result in different negativities. This is so, since entanglement is dependent on operator ordering  as we showed above. However, it is also true that even if we do not interchange these operators, and only permute $c^\dagger$ with any one of them, entanglement on the reduced state changes. This is remarkable because $c^\dagger$ disappears after taking the partial trace. As a matter of fact, if we take the ordering $(a^\dagger c^\dagger b^\dagger)$, so as to have $\ket{111}=a^\dagger c^\dagger b^\dagger\ket{0}$, the last term in \eq{emix} changes sign and the reduced state is
\begin{align}\rho=\frac{1}{2}\left(\proj{10}{10}+\proj{01}{01}\right)\end{align}
which is a completely unentangled state. The negativity is thus zero.

A situation of the kind described above appears in the context of quantum field theory in non-inertial frames and curved spacetimes; see section \ref{main} and \cite{AlsingSchul,Edu2}. Specifically, in certain cases (uniformly accelerated observers, entanglement in the presence of black holes, etc.), there are regions of spacetime which are causally disconnected from the observer's world-line. Therefore, modes living inside them are not allowed to communicate with the observer. One is therefore forced to trace them out \cite{Alicefalls,AlsingSchul}. In the context of fermionic fields, the relative positions of these modes will affect entanglement, as above. Thus the entanglement in the field state depends on the sign criterium over modes which do not have a causal connection with the observer. 

It is obvious that physical observations cannot be changed by an operator ordering criterium. Indeed, a specific ordering is imposed by the physical nature of experiments as we show in section \ref{ord}. However, this fact has been overlooked in previous studies in relativistic quantum information, assuming somehow that an arbitrary convention can be taken to derive general results. Although this is true in some cases, we will show that remarkable differences appear when we introduce a physical criterion to select the basis.

We conclude this section with the remark that there does not seem to be any way to construct an ordering-independent entanglement measure for fermionic states. Any reasonable entanglement measure should vanish for separable states and achieve its maximum at maximally entangled states. But as our examples show, separable states in one basis need not be separable in another, and can in fact be maximally entangled. It is therefore impossible to define a reasonable entanglement measure for these states.

\section{Operator ordering in relativistic quantum field theory}\label{main}

The effects of operator ordering will now be discussed for the kind of states that appear in relativistic quantum information. Specifically, we will consider  a 1+1 flat spacetime in which there is an observer, Rob, who undergoes uniform proper acceleration, and another observer, Alice, who stays inertial. Both look at the same state of a fermionic field of spin $s$, but while Alice uses a basis of Minkowski modes (plane waves in the massless case) for the description of her part of the field state, Rob uses the so-called Rindler modes \cite{Edu9}. These modes are the natural candidates to describe the state of a quantum field from the viewpoint of an accelerated observer, since a uniformly accelerated detector would couple to them. Due to the nature of the change of basis between Rindler and Minkowski modes (given by the so-called Bogoliubov coefficients; see \cite{Jauregui,Langlois,Edu9}), which mixes  Minkowski creation and annihilation operators, the Minkowski and Rindler vacuums are not the same. This is the origin of the celebrated Unruh effect \cite{Unruh0}.

\begin{figure}[hbtp] 
\includegraphics[width=.50\textwidth]{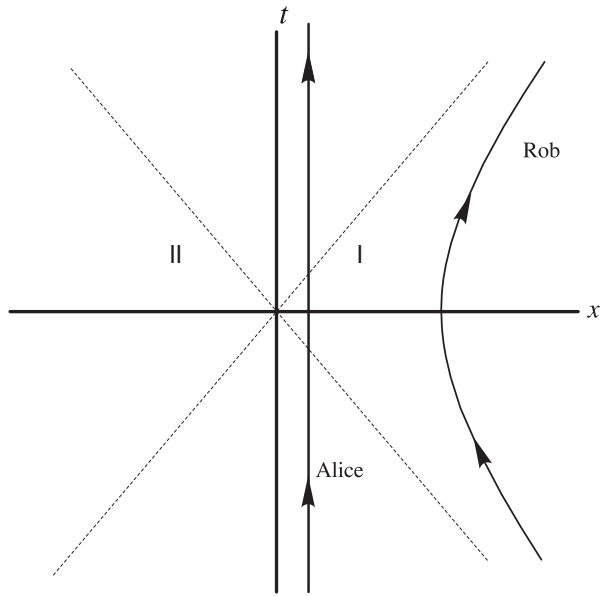}
\caption{Minkowski spacetime diagram showing the world lines of an inertial observer Alice, and one uniformly accelerated observer moving hyperbolically in region I. Note that regions I and II are causally disconnected from each other.}
\label{rindler}
\end{figure}

Rindler modes are named after Rindler coordinates $(\eta,\tau)$, which describe a family of uniformly accelerated observers in two causally disconnected patches of spacetime, labelled as regions I and II in Figure \ref{rindler}. The observers are right-moving in region I and left-moving in region II. An observer following an orbit of constant $\eta$ is uniformly accelerated and sees the boundary of region I as an event horizon. Likewise, the same happens for an observer following an orbit of constant $\eta$ in region II. Rindler modes are monochromatic solutions of the field equation in Rindler coordinates, therefore they only have support inside either region I or II. This means that for each particle species, frequency $\omega$ and spin-z component we have two annihilation operators, one for each region. The operator $c_{\omega,s,\text{I}}$ will correspond to the Rindler mode of frequency $\omega$ and spin $\sigma$ in region I, and $d_{\omega,s,\text{I}}$ will correspond to its antiparticle, with the same considerations applying to region II. 

The existence of these two kinds of modes implies that, from the accelerated observer viewpoint, the Hilbert space factorises as $\mathcal{H}_\text{I}\otimes \mathcal{H}_\text{II}$. Any physical accelerated observer such as Rob must remain in either region I or II, since both are causally disconnected. In quantum-mechanical terms, this means that we must trace out the part of the state outside the observer's region, since it will not be observed. Customarily we take the observer Rob to be in region I, and so henceforth we will trace out $\mathcal{H}_\text{II}$.

There is another set of modes which can be constructed by taking simple linear combinations of Rindler modes \cite{Edu9},
\begin{align}C^\dagger_{\omega,\sigma,\text{R}}&=\cos r_\omega c^\dagger_{\omega,\sigma,\text{I}}- \sin r_\omega d_{\omega,-\sigma,\text{II}},\nonumber\\
C^\dagger_{\omega,\sigma,\text{L}}&=\cos r_\omega c^\dagger_{\omega,\sigma,\text{II}}- \sin r_\omega d_{\omega,-\sigma,\text{I}},\label{umodes}\end{align}
where $\tan r_\omega=e^{-\pi\omega c/a}$ and $\omega$ indicates Rindler frequency. These are the so-called Unruh modes. The subscripts L and R stand for `left' and `right' modes, which are related to each other by a reversal of regions I and II. These modes have the particularity of being linear combinations of purely positive-frequency Minkowski modes. In terms of creation and annihilation operators, this means that \eq{umodes} can be rewritten as a linear combination of only Minkowski creation operators. This trivially implies that the Minkowski and Unruh vacua are the same. The Minkowski vacuum can then be written as a product of the vacua for each Unruh mode,
\begin{align}\ket{0}_\text{M}=\bigotimes_{\omega}\ket{0}_{\omega,\text{U}}\label{product}\end{align}
where each $\ket{0}_\text{U}$ is annihilated by $C_{\omega,\sigma,\text{R}}$ and $C_{\omega,\sigma,\text{L}}$ for all $\sigma$. 

In this  work (as in previous literature \cite{Alicefalls,AlsingSchul}), the accelerated observer Rob watches a single Unruh mode of the field. Given \eq{product}, we only need to perform a detailed study of one of the factors in the tensor product, and hence from now on we shall drop the frequency index in all operators. We will consider arbitrary Unruh excitations instead of restricting ourselves to the so-called single mode approximation \cite{Edu9}
\begin{align}\ket{1}_{\sigma,\text{U}}&=C^\dagger_{\omega,s,\text{U}}\ket{0}_\text{U}=\left(\qr C^\dagger_{\omega,\sigma,\text{R}}+\ql C^\dagger_{\omega,\sigma,\text{L}}\right)\ket{0}_\text{U},\nonumber\\ &\quad \vert\qr\vert^2+\vert\ql\vert^2=1,\end{align}
The case $\qr=1$ corresponds to the choice previously known as the single mode approximation \cite{Alsingtelep,AlsingMcmhMil}.
 
Now, the Unruh vacuum and excitations are expressed in a simple way in the Rindler basis, due to the trivial form of \eq{umodes}. Note that no considerations whatsoever of operator ordering have been assumed so far: To fix a convention, we will assume Alice's mode appears leftmost in all expressions. It is also assumed that the mode watched by the inertial observer Alice possesses a negligible overlap with the Unruh mode, so that we can consider both subsystems as noninteracting.

For concreteness, we will be interested in states of the form
\begin{align}\ket{\Psi}&=P\ket{0}_\text{A} (A^\dagger_\text{U}\ket{0}_\text{U})+Q\ket{1}_\text{A} (B^\dagger_\text{U}\ket{0}_\text{U}),\label{est}\\\nonumber  &\quad \vert P\vert^2+\vert Q\vert^2=1,\end{align}
where $A_\text{U}$ and $B_\text{U}$ are arbitrary linear combinations of products of Unruh modes $C_{\sigma,\text{U}}$ so that Rob's state is in general not a qubit. Note that the second part of the state will have to be expressed in the Rindler basis.

In order to express $\ket{0}_\text{U}$ in the Rindler basis, we will factor the vacuum as $\ket{0}_\text{U}=\ket{0}_\text{R}\otimes\ket{0}_\text{L}$ with $C_{\sigma,\text{R}}\ket{0}_\text{R}=0$ and $C_{\sigma,\text{L}}\ket{0}_\text{L}=0$ for all $\sigma$ as in \cite{Edu9}. This factorisation already entails an assumption on the ordering of the Rindler operators. Namely, the operators $c^\dagger_{\omega,\sigma,\text{I}}$ and $d^\dagger_{\omega,-\sigma,\text{II}}$ should always appear to the left of the operators $c^\dagger_{\omega,\sigma,\text{II}}$ and $d^\dagger_{\omega,-\sigma,\text{I}}$. 

We will now choose a specific ordering for both the left and right sectors as follows:  within each sector, all region I operators shall appear before region II operators and, within each region, operators will be ordered by their spin-z component, with the highest value appearing leftmost. We can associate a binary number to each element of the basis. In this fashion, the vacuum state for each sector would be indexed by a sequence of $2\cdot(2s+1)$ zeros. The one-particle excitations would have a 1 in the $k$th position, and so on.  For example, for $s=\tfrac{1}{2}$, we would have
\begin{align}\ket{1111}_{\text{R}}=c^\dagger_{\up,\text{I}}c^\dagger_{\down,\text{I}}d^\dagger_{\up,\text{II}}d^\dagger_{\down,\text{II}}\ket{0}_\text{Rindler},\nonumber\\ \ket{1111}_{\text{L}}=d^\dagger_{\up,\text{I}}d^\dagger_{\down,\text{I}}c^\dagger_{\up,\text{II}}c^\dagger_{\down,\text{II}}\ket{0}_\text{Rindler}.\end{align}

This notation coincides with that used for the Grassman field in previous works \cite{Edu9} and  has the advantage that it is easy to find and write the vacuum state for arbitrary spin, as we will show below. 

In the literature for spin $1/2$ fields, \cite{Edu2} the common notation groups the spin-up and spin-down operators for each region and particle species. To translate from the generic notations introduced above to the short one for Dirac fields, we take the indices in groups of two and make the replacements $00\rightarrow0$, $01\rightarrow\down$, $10\rightarrow\up$, and $11\rightarrow p$. The $p$ stands for `pair'. For instance, we have
\begin{align}\ket{1011}_\text{R}\ket{0110}_\text{L}=\ket{\up\! p\!\down\up}\nonumber\end{align}

We now can calculate the right vacuum by assuming the following ansatz, in which again we have introduced new notation:
\begin{align}\ket{0}_\text{R}=\sum_{\alpha}x_\alpha\ket{\alpha\; \mathcal{R}(\alpha)}_\text{R}.\label{vansatz}\end{align}
Here, $\alpha$ is a binary number with length $2s+1$, and $\mathcal{R}(\alpha)$ is the binary number obtained by reversing the ordering of the figures of $\alpha$ [e.g. if $\alpha=01\Rightarrow \mathcal{R}(\alpha)=10$]. The sum is extended to all $2^{2s+1}$ such chains. The chain $\alpha \mathcal{R}(\alpha)$ is obtained by concatenating them so that $\alpha$ is associated with modes in region I and $\mathcal{R}(\alpha)$ with modes in region II. The ansatz \eq{vansatz} is of the form of a squeezed vacuum state.

We shall define a few operations and notation on these chains: $\chi(\alpha,k)$ will denote the number of 1's in $\alpha$ before its $k$th position. $\chi(\alpha)$ will simply be the sum of all the digits in $\alpha$. If $\alpha$ is a chain with a 0 in its $(2s+2-k)$th position, then $\alpha+2^k$ will be the chain with a 1 in its $k$th position and the same digits as $\alpha$ elsewhere, mimicking the addition of binary numbers. Finally, $S_k$ will be the set of all $\alpha$ with a 0 in its $(2s+2-k)$th position.

In order to determine the coefficients $x_\alpha$ we impose $C_{\sigma,\text{R}}\ket{0}_\text{R}=0$ for all $\sigma$ and therefore get
\begin{align}0&=\sum_\alpha x_\alpha\!\!\left[\cos\ro c_{\sigma,\text{I}}\ket{\alpha\; \mathcal{R}(\alpha)}- x_\alpha\sin\ro d^\dagger_{-\sigma,\text{II}}\right]\!\ket{\alpha\; \mathcal{R}(\alpha)}_\text{R}\nonumber\\
&=\sum_{\alpha\not\in S_\sigma} x_\alpha \cos\ro  c_{\sigma,\text{I}}\ket{\alpha\; \mathcal{R}(\alpha)}_\text{R} \nonumber\\&-\sum_{\alpha\in S_\sigma} x_\alpha\sin\ro d^\dagger_{-\sigma,\text{II}}\ket{\alpha\; \mathcal{R}(\alpha)}_\text{R}\nonumber\\&=\sum_{\alpha\in S_\sigma} (-1)^{\chi(\alpha,\sigma)}  x_{\alpha+2^\sigma}\cos\ro \ket{\alpha\; \mathcal{R}(\alpha+2^\sigma)}_\text{R} \nonumber\\&- (-1)^{\chi(\alpha)+\chi(\mathcal{R}(\alpha),2s+2-\sigma)} x_\alpha\sin\ro \ket{\alpha\; \mathcal{R}(\alpha+2^\sigma)}_\text{R}\nonumber\\&=\sum_{\alpha\in S_\sigma}(-1)^{\chi(\alpha,\sigma)}\left(x_{\alpha+2^\sigma}\cos\ro-x_\alpha\sin\ro\right) \nonumber\\&\;\quad \ket{\alpha\; \mathcal{R}(\alpha+2^\sigma)}_\text{R}\end{align}
In passing to the last line, we used the property $\chi(\mathcal{R}(\alpha),2s+2-\sigma)=\chi(\alpha)-\chi(\alpha,\sigma)$ for all $\alpha\in S_\sigma$. By equating each coefficient to zero we obtain the recurrence relations
\begin{align}x_{\alpha+2^k}=\tan\ro\cdot x_\alpha,\end{align}
which can be solved iteratively by taking, e.g, $x_{0\ldots0}=1$. Then, $x_\alpha=\tan\ro^{\chi(\alpha)}$, and upon imposing normalisation, we obtain for the right vacuum
\begin{align}\ket{0}_\text{R}=\sum_{\alpha}(\cos\ro)^{2s+1-\chi(\alpha)}(\sin\ro)^{\chi(\alpha)}\ket{\alpha\; \mathcal{R}(\alpha)}_\text{R}\label{vr}\end{align}
The left vacuum is straightforward to obtain from \eq{vr} by noting that the ordering in the left sector interchanges particles and antiparticles. The left vacuum therefore is
\begin{align}\ket{0}_\text{L}\!\!=\!\!\sum_{\alpha}(-1)^{\chi(\alpha)}(\cos\ro)^{2s+1-\chi(\alpha)}(\sin\ro)^{\chi(\alpha)}\ket{\alpha\, \mathcal{R}(\alpha)}_\text{L}\label{vl}\end{align}
We remark again that in deriving \eq{vr} and \eq{vl} we relied on a very specific operator ordering. As seen in section \ref{toy}, entanglement measures will not remain invariant in general when this ordering is changed. This is a point so far neglected in the literature.%PRA: This point reveals an ambiguity in previous results on fermionic field entanglement in non- inertial frames.
We will now study the changes in negativity for the state \eq{est} and different values of $s$.
\subsection{Grassmann field}
We shall now consider the behaviour of entanglement under different operator orderings in the Grassman scalar field, which is an anticommuting scalar field with only one degree of freedom. This field has been extremely useful to study the general features of entanglement in fermionic fields \cite{AlsingSchul,chapucilla,chapucilla2,Shapoor,Geneferm,chor1,Edu9,chor2}. 

Under these circumstances, we can use \eq{vr} and \eq{vl} together with $\ket{0}_\text{U}=\ket{0}_\text{R}\otimes\ket{0}_\text{L}$ to obtain the Unruh vacuum in terms of the Rindler vacuum, using the operator ordering  $c^\dagger_\text{I}\ d^\dagger_\text{II}\ d^\dagger_\text{I}\ c^\dagger_\text{II}$:
\begin{align}\label{gvac}\ket{0}_\text{U}&=\cos^2\ro\ket{0000}-\sin\ro\cos\ro\ket{0011}\nonumber\\&+\sin\ro\cos\ro\ket{1100}-\sin^2\ro\ket{1111}.\end{align}
The one-particle excitations are obtained as
\begin{align}\label{gexc}\ket{1}_\text{U}&=(\qr C^\dagger_{\text{R}}+\ql C^\dagger_{\text{L}})\ket{0_\text{U}}\nonumber\\&=\qr\left[\cos\ro\ket{1000}-\sin\ro\ket{1011}\right]\nonumber\\&+\ql\left[\sin\ro\ket{1101}+\cos\ro\ket{0001}\right]\end{align}
and the general state \eq{est} takes the simple form
\begin{align}\ket{\Psi}&=P\ket{0}_\text{A}\left[a_1\ket{0}_\text{U}+b_1\ket{1}_\text{U}\right]+Q\ket{1}_\text{A}\left[a_2\ket{0}_\text{U}\right.\nonumber\\&\left.+b_2\ket{1}_\text{U}\right],\quad\quad \vert a_i\vert^2+\vert b_i\vert^2=1\; \text{for}\ i=1,2.\label{gest}\end{align}

We now compute the density matrix for the state, and then trace out region II operators for the reasons described above. Finally we compute the negativity as an entanglement measure for the Alice-Rob system. For a general state and values of $\qr$, entanglement can be created as shown in \cite{Mig1}, in contrast with the common conception that the Unruh effect should effectively act as a thermal bath.

However, we could do this in a different operator ordering. For instance, we could rearrange the region II operators in any way we want while leaving the relative positions of the Alice mode operator and Rob's region I operators unchanged. A naive expectation would be that since region II modes live in a different causal patch of spacetime, their position should be irrelevant when considering entanglement.

\begin{figure}[hbtp] 
\includegraphics[width=.50\textwidth]{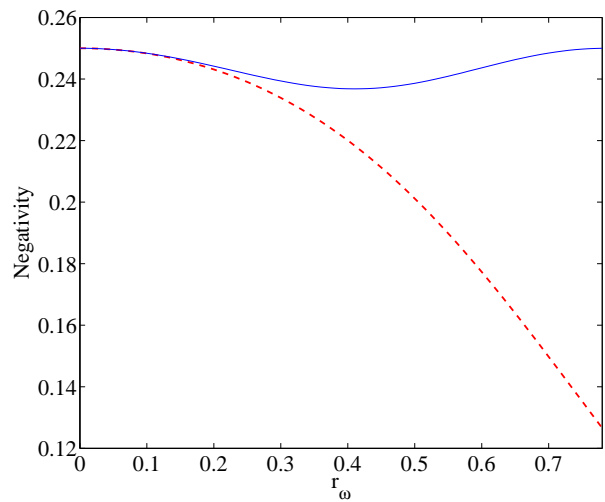}
\caption{(Colour online) Negativity as a function of $\ro$ for $\qr=1/\sqrt{2}$, the state \eq{gest} with $P=1/\sqrt{2}$, $a_2=b_1=0$, $a_1=b_2=1$ and operator orderings $c^\dagger_\text{I}\ d^\dagger_\text{II}\ d^\dagger_\text{I}\ c^\dagger_\text{II}$ (red dashed curve) and $c^\dagger_\text{I}\ d^\dagger_\text{I}\ d^\dagger_\text{II}\ c^\dagger_\text{II}$ (blue solid curve). Note that in the first case, entanglement is a monotonically decreasing function of $\ro$ and thus acceleration, whereas in the second, entanglement is generated for $\ro\geq\tfrac{\pi}{8}$.}
\label{figgras}
\end{figure}

Nevertheless, this is not the case. The expression for the state \eq{gest} in any operator ordering is readily obtained from \eq{gvac}, \eq{gexc}, and \eq{gest} by permuting the operators in each term of the superposition, and taking into account the corresponding signs. Then, when region II modes are traced, the phenomenon discussed at the end of section \ref{toy} appears: There are indeed different negativity behaviours corresponding to different operator orderings, as shown in Fig.\ref{figgras}. Note that in the Grassman case, only two different behaviours appear for a general state. This can be traced back to the fact that by performing some change of basis we can get rid of all the signs in the partial transpose of the reduced density matrix, except for one.

There is always some negativity that survives in the infinite acceleration limit, in concordance with previous results \cite{AlsingSchul,Edu2,Edu3,Edu6,Chinos,Edu9,Edu10}. This is true for all the states and fermionic fields we have considered. The survival of entanglement at infinite acceleration is therefore a physical phenomenon and not an artifact of choosing a specific ordering. Moreover, there is evidence that the phenomenon is related to a tradeoff of entanglement between particle and antiparticle sectors \cite{Edu10}. 

Remarkably, the state shows entanglement regeneration in one ordering which is absent in the other. Our results in section \ref{toy} imply that any other entanglement measure will present this ambiguity. What is needed is a physical criterion for choosing a specific ordering; see section \ref{ord}.

Finally, we note that if we restrict ourselves to $\qr=1$, the case previously known in the literature as the single mode approximation, there is only one negativity behaviour. This is a quirk of the Grassman field and it is not true for general spin, as we shall see below.

\subsection{Dirac field}
The Dirac field presents a much richer zoo of negativity behaviours, as could be expected from the doubling in the number of Rindler operators from the Grassman field. The general state \eq{est} now takes the form
\begin{align}\ket{\Psi}&=P\ket{0}_\text{A}\left[a_1\ket{0}_\text{U}+b_1\ket{\up}_\text{U}+c_1\ket{\down}_\text{U}+d_1\ket{p}_\text{U}\right]\nonumber\\&+Q\ket{1}_\text{A}\left[a_2\ket{0}_\text{U}+b_2\ket{\up}_\text{U}+c_2\ket{\down}_\text{U}+d_2\ket{p}_\text{U}\right],\label{dest}\end{align}
where
\begin{equation}
\vert a_i\vert^2+\vert b_i\vert^2+ \vert c_i\vert^2+ \vert d_i\vert^2=1\; \text{for}\ i=1,2,
\end{equation}
and we have defined $\ket{p}_\text{U}=C^\dagger_{\up,\text{U}}C^\dagger_{\down,\text{U}}\ket{0}_\text{U}$. The vacuum and excitations can be computed from \eq{vr} and \eq{vl} in the same way as in the Grassman scalar case. The implicit ordering is  $c^\dagger_{\up,\text{I}}\ c^\dagger_{\down,\text{I}}\ d^\dagger_{\up,\text{II}}\ d^\dagger_{\down,\text{II}}\ d^\dagger_{\up,\text{I}}\ d^\dagger_{\down,\text{I}}\ c^\dagger_{\up,\text{II}}\ c^\dagger_{\down,\text{II}}$. In order to make the results more readable, we shall use the abbreviated notation for the Dirac field defined above. The translation from the notation in which \eq{vr} and \eq{vl} are written to the new one is straightforward. With these conventions, the vacuum, excitation, and pair terms are
\begin{align}\nonumber\ket{0}_\text{U} &= \cos^4\ro\ket{0000}-\cos^3\ro\sin\ro(\ket{00\!\up\down}+\ket{00\!\down\up}) \\\nonumber &+ \cos^2\ro\sin^2\ro\ket{00pp} \\\nonumber &+\cos^3\ro\sin\ro(\ket{\up\down\!00} + \ket{\down\up\!00}) \\\nonumber &-\sin^2\ro\cos^2\ro(\ket{\up\down\up\down}+\ket{\up\down\down\up}+\ket{\down\up\up\down}+\ket{\down\up\down\up})\\\nonumber &+\cos\ro\sin^3\ro(\ket{\up\down\! pp}+\ket{\down\up\! pp})+\sin^4\ro\ket{pppp}\\\nonumber &-\cos\ro\sin^3\ro(\ket{pp\!\up\down}+\ket{pp\!\down\up})\\&+\cos^2\ro\sin^2\ro\ket{pp00},\label{dvac}\end{align}
\begin{align}
\nonumber\ket{\sigma}_\text{U} &= q_\text{L}[\cos^3\ro\ket{000\sigma}\!+\cos^2\ro\sin\ro(\ket{\up\down\!0\sigma}+\ket{\down\up\!0\sigma})\\\nonumber &+ \cos\ro\sin^2\ro\ket{pp0\sigma}) \\\nonumber &+\sgn (\sigma)\cdot(\cos^2\ro\sin\ro\ket{00\sigma p}\\\nonumber &+ \cos\ro\sin^2\ro(\ket{\up\down\!\sigma p}+\ket{\down\up\!\sigma p})+\sin^3\ro\ket{pp\sigma p})]\\\nonumber &+ \qr[\cos^3\ro\ket{\sigma000}- \cos^2\ro\!\sin\ro (\ket{\sigma0\!\up\down}+\ket{\sigma0\!\down\up}) \\\nonumber &+\cos\ro\sin^2\ro\ket{\sigma 0pp} \\\nonumber &+ \sgn (\sigma)\cdot(\cos^2\ro\sin\ro\ket{p\sigma00}\\ & +\sin^3\ro\ket{p\sigma pp}- \cos\ro\sin^2\ro(\ket{p\sigma\!\up\down}+\ket{p\sigma\!\down\up}))],\label{dexc}\end{align}
\begin{align}
\ket{p}_\text{U}&=\qr^2\left(\cos^2\ro\ket{p000}-\sin\ro\cos\ro(\ket{p0\!\up\down}+\ket{p0\!\down\up})\right.\nonumber\\&\left.+\sin^2\ro\ket{p0pp}\right)\nonumber\\&+\ql^2\left(\cos^2\ro\ket{000p}+\sin\ro\cos\ro(\ket{\up\down\!0p}+\ket{\down\up\!0p})\right.\nonumber\\&\left.+\sin^2\ro\ket{pp0p}\right)
\nonumber\\&+\qr\ql\left(\cos^2\ro\ket{\up\!00\!\down}-\cos\ro\sin\ro\ket{\up\!0\!\down\! p}\right.
\nonumber\\&+\sin\ro\cos\ro\ket{p\!\up\!0\!\down}-\sin^2\ro\ket{p\!\up\!\down\! p}
\nonumber\\&-\cos^2\ro\ket{\down\!00\!\up}-\cos\ro\sin\ro\ket{\down\!0\!\up\! p}
\nonumber\\&\left.+\sin\ro\cos\ro\ket{p\!\down\!0\!\up}+\sin^2\ro\ket{p\!\down\!\up\! p}\right).\label{dpar}
\end{align}

Using \eq{dvac}, \eq{dexc}, and \eq{dpar}, we can study negativity for an arbitrary state of the form \eq{dest} and operator ordering, by making adequate permutations. It is not straightforward to compute explicitly how many negativity behaviours can arise, since there are $8!=40320$ possible orderings. Many of these are equivalent regarding entanglement: For instance, any permutation which involves only transpositions of operators of the same region constitutes a local unitary and thus leaves entanglement unchanged. For some special states, there can even be more symmetry: If $\qr=1$, then all the `left' operators are irrelevant, meaning that their position in the ordering does not affect negativity. The converse is true when $\ql=1$. If the state only carries spin-up excitations, all the spin-down particle operators and spin-up antiparticle operators are irrelevant. 

\begin{figure}[hbtp] 
\includegraphics[width=.50\textwidth]{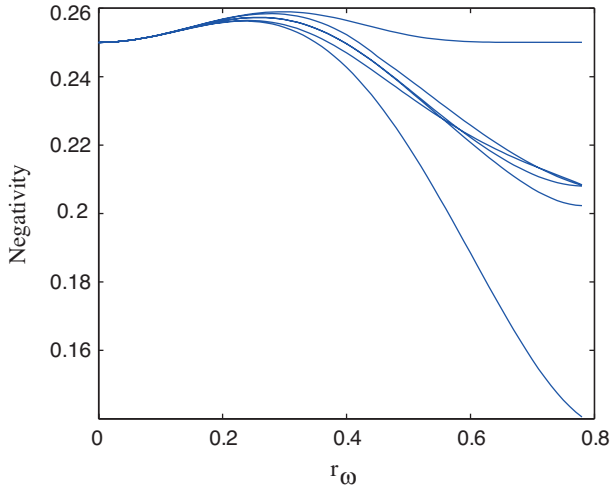}
\caption{Negativity as a function of $\ro$ for the Dirac singlet state when $\qr=1/\sqrt{2}$ and different operator orderings. The general behaviour is similar in all such cases: There is a slight entanglement creation which is then destroyed. For every ordering, some entanglement survives at $\ro=\tfrac{\pi}{4}$.}
\label{dirsing}
\end{figure}

Nevertheless, a survey of all the $8!$ orderings was carried out numerically. The results vary wildly depending on the degree of symmetry of the state. For the singlet state, \eq{dest} with $a_1=c_1=d_1=a_2=b_2=d_2=0$, $b_1=c_2=1$ and $P=1/\sqrt{2}$ there are only six negativity behaviours, as depicted in fig \ref{dirsing}. At first negativity rises steadily, then reaches a maximum after which it decreases monotonically. For the canonical ordering implicit throughout this work, negativity tends to its inertial value of $\tfrac{1}{4}$ from above as $\ro\rightarrow\infty$, never decreasing past this value. For the ordering $c^\dagger_{\up,\text{I}}\ d^\dagger_{\down,\text{II}}\ d^\dagger_{\down,\text{I}}\ c^\dagger_{\up,\text{II}}\ c^\dagger_{\down,\text{I}}\ d^\dagger_{\up,\text{II}}\ d^\dagger_{\up,\text{I}}\ c^\dagger_{\down,\text{II}}$ negativity decreases far more quickly, reaching only $0.1398$ in the infinite acceleration limit. All the other behaviours are intermediate between these two, and are qualitatively similar to the second.
The six classes are not evenly populated: The uppermost one contains 4032 different operator orderings and is the least abundant, while the bottommost one contains 9408 operator orderings and is the most abundant.

If we consider a general state excluding pair terms, that is, a state of the form \eq{dest} with $d_1=d_2=0$, the number of negativity classes rises to 64, which suggests  that, in analogy with the Grassman case, the choice between one behaviour and another is governed by six signs. This is, however, disproved by the differing number of orderings which show each behaviour: If the choice depended only on six signs, then all classes should be equally populated. We also note that for $\qr=1$, the 64 behaviours collapse into only two, which is not surprising given the high degree of symmetry present in this case.

Finally, considering \eq{dest} with arbitrary coefficients results in 778 classes, which lack any exploitable structure. There are two effects determining the number and population of the classes: On one side, we have the different signs which can arise for each term of the superposition in \eq{dest} for each operator ordering. On the other, the density matrix (and thus the partial transpose used in computing negativity) is zero in most of its entries. This makes some sign changes  trivial,  which could otherwise have resulted in a different negativity. 

\subsection{Higher spin}
A similar treatment to those of the Grassman and Dirac fields can be performed for fermionic fields of half-integer spin. The complexity of the task grows quickly: There are $[4\cdot(2s+1)]!$ possible orderings for a field of spin $s$. This means that already for $s=\tfrac{3}{2}$, for which there are $2.1\cdot10^{13}$ orderings, the use of  Monte Carlo algorithms is essential. 

\begin{figure}[hbtp] 
\includegraphics[width=.50\textwidth]{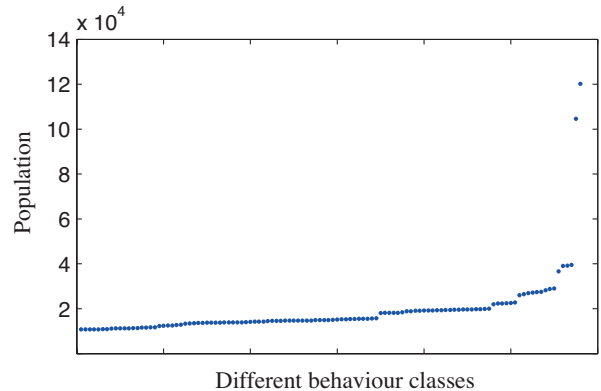}
\caption{Population for some negativity classes in the $s=\tfrac{3}{2}$ case, including the most abundant ones. Note that the points group in `steps' of classes with the same negativity, and the two dominant classes are well above the others.}
\label{hist}
\end{figure}

We considered the particular instance of  \eq{est} given by $P=1/\sqrt{2}$ and
\begin{align}A_\text{U}=C_{+\tfrac{3}{2},\text{U}}+C_{+\tfrac{1}{2},\text{U}},\quad B_\text{U}=C_{-\tfrac{1}{2},\text{U}}+C_{-\tfrac{3}{2},\text{U}}\end{align}
This state is equivalent to a Dirac singlet state by making a simple change of spin basis, but the number of negativity behaviours is much higher because of the greater number of operators (and thus possible permutations). We found $\approx1.4\cdot10^6$ different behaviours for negativity. As can be seen in Fig.\ref{hist}, there are two dominant behaviours much more frequent than the others, and a plethora of infrequent behaviours. There are also families of different negativity behaviours with the same population. The natural explanation for these phenomena is that each of these families correspond to a set of orderings with the same symmetries.

\section{Choosing a physical ordering}\label{ord}
We have seen in section \ref{main} that there can be an enormous number of different negativity behaviours for each different operator ordering that shuffles region II operators. This leaves us in an uncomfortable position, since operator orderings are purely conventional and therefore \emph{a priori} equivalent. But if all the orderings seem to be on an equal footing, it is only because we are lacking a physical criterion for which class of them should be chosen. We will find such a  criterion in this section.

First, note that all observables such as expected values or transition probabilities are obviously manifestly independent of operator ordering: One can always use the anticommutation relations to transform any matrix element into a superposition of terms proportional to $\braket{0}{0}=1$, and this transformation is unique. Although negativity is not an observable, it can be experimentally determined (for instance by means of state tomography), which means that it ought to be expressible in terms of expectation values. Nevertheless, it is not the negativity of the field that which can be measured, but that of the detector. Indeed, for any particular detector-field model, we look at the state of the coupled system, usually long after the interaction has been turned off, and then trace out the `field' subsystem. The negativity of the detector's entangled state can then be experimentally determined. A field is never really observed; it interacts with a detector which is the only experimentally accessible system.

This means that in order to determine a suitable ordering we must study the form of the interaction between the detector and the field. A physical detector for fermionic fields has not yet been proposed. Nevertheless, there is an obvious causality condition that any physical detector should fulfill: The interaction Hamiltonian for a detector moving in region I of Rindler spacetime cannot contain any region II operators, and vice versa. 

We seek to find the conditions in which all of the field entanglement is physically meaningful, i.e., it can be observed. Without a fermionic detector model at our disposal, we cannot address this question directly; it may well be the case that no physical detector can ever acquire all of the field correlations studied here and elsewhere. In this case, we would be justified in labeling this field entanglement as unphysical.

However, it may be the case that some particular detector can be imprinted with all of the field entanglement. If so, there will be a particular operator ordering, dependent on the exact nature of the field-detector interaction, in which no anticommutation signs will arise in any computation (simply take the operator ordering to be the reverse of the fermionic operator ordering in the interaction Hamiltonian). Since the interaction only has region I fermionic operators, any such ordering must have all region II operators rightmost. Permutation of the operators of the same region constitutes a local unitary transformation, which does not change entanglement, and therefore any two such   orderings will show the same entanglement.

We therefore have found a necessary condition for field entanglement to be physical. This condition selects one and only one entanglement behaviour class and therefore settles the question of what ordering should be chosen to compute fermionic entanglement. We also point out that the resulting negativity in states such as \eq{gest} is different from the one reported in \cite{Edu9}, which corresponds to a different class of operator orderings. Namely, these works found the monotonically decreasing red dashed curve in Fig. \ref{figgras}, whereas the physically meaningful curve is the blue solid one, which shows a completely different behaviour.

To the authors' knowledge, the only physical system that both couples directly to fermion fields and is easily measurable, the electromagnetic field, does so through an interaction quadratic on the field. The detector models studied in \cite{Takagi,Kumar}  considered this kind of quadratic couplings. With some additional structure, these may well constitute physical models for a fermionic detector. However, these simplified detectors would not be able to acquire all the entanglement from the fermionic field modes, save for a few very specific states.

Of course, all the considerations made above apply to this fermionic detector as well as to any other, even though the detector is not capable of measuring field entanglement completely.

\section{Conclusions}\label{cojoniak}

In this paper we tackle an ambiguity present in all previous works on fermionic entanglement in non-inertial frames. The amount of entanglement is not invariant under nonlocal changes of basis. This is a general feature of quantum systems and hardly new.

When working with fermionic systems as support for qubits, we must be careful, since naively defining the basis of the multipartite Hilbert space without paying attention to the sign convention may lead to unphysical results. We have shown that an entanglement measure can behave completely differently if we consider the different bases naturally suggested by the fermionic operator orderings.

This situation has a special relevance in the context of fermionic fields in non-inertial frames, where part of the system has to be traced out. It could then be naturally expected that once those modes have been traced out, they would not affect entanglement of the remaining system. Nevertheless, we have seen that this intuition is not true.

Of course, physics cannot depend on a mathematical convention. We have shown how when we introduce detectors, which are a fundamental requirement to observe the field, the ambiguity disappears. Field entanglement is only physical if it can be acknowledged by detectors coupled to the field. There is only one set of bases (all of which share the same entanglement properties) for which field entanglement can be imprinted on detectors. We have shown the form of a simple detector model (consisting of an array of Unruh-Dewitt-like detectors) for which all the entanglement can be transmitted from the field to the detector.

Incidentally, entanglement in this scenario happens to behave in a different way from that for the bases considered in previous works in the literature on fermionic entanglement in non-inertial frames \cite{Edu2,Edu9,Edu10}.

\section{Acknowledgments}

We thank Juan Le\'on for our interesting and useful discussions related to this work. We also thank Jorma Louko for his very helpful comments.  

Eduardo Mart\'in-Mart\'inez was supported by a CSIC JAE-PREDOC2007 Grant, the Spanish MICINN Project FIS2008-05705/FIS, and the QUITEMAD consortium.

%if any ordering in which region II operators appear to the left of region I operators (unless there are region I operators which do not appear in the expression of the interaction Hamiltonian in terms of creation and annihilation operators; such field modes are not seen by the detector anyway)
\end{document}